\documentclass{article}

\usepackage{PRIMEarxiv}
\usepackage{amsmath} 
\usepackage{xcolor}
\usepackage{graphicx}
\usepackage{caption}

\usepackage{amssymb}
\usepackage{geometry}
\usepackage{array}
\usepackage{xcolor}
\usepackage{pifont}
\usepackage{longtable}
\usepackage{textcomp}
\usepackage{listings}

\usepackage[utf8]{inputenc} 
\usepackage[T1]{fontenc}    
\usepackage{hyperref}       
\usepackage{url}            
\usepackage{booktabs}       
\usepackage{amsfonts}       
\usepackage{nicefrac}       
\usepackage{microtype}      
\usepackage{lipsum}
\usepackage{fancyhdr}       
\usepackage{graphicx}       
\graphicspath{{media/}}     

\newcommand{\cmark}{\ding{51}} 
\newcommand{\xmark}{\ding{55}} 

\title{LLM Agents Improve Semantic Code Search  
}

\author{
  Sarthak Jain \\
  Cisco \\
  University of Illinois Urbana Champaign \\
  Urbana, Illinois \\
  \texttt{sj84@illinois.edu, sarjain2@cisco.com} \\
   \And
  Aditya Dora \\
  University of Illinois Urbana Champaign \\
  Urbana, Illinois \\
  \texttt{adora2@illinois.edu} \\
   \And
   Ka Seng Sam \\
   University of Illinois Urbana Champaign \\
   Urbana, Illinois \\
   \texttt{samsam2@illinois.edu}
   \And
  Prabhat Singh \\
  Cisco \\
  San Jose, California\\
  \texttt{prabhat7@cisco.com} \\
}

\begin{document}
\maketitle

\begin{abstract}
Code Search is a key task that many programmers often have to perform while developing solutions to problems. Current methodologies suffer from an inability to perform accurately on prompts that contain some ambiguity or ones that require additional context relative to a code-base. We introduce the approach of using Retrieval Augmented Generation (RAG) powered agents to inject information into user prompts allowing for better inputs into embedding models. By utilizing RAG, agents enhance user queries with relevant details from GitHub repositories, making them more informative and contextually aligned. Additionally, we introduce a multi-stream ensemble approach which when paired with agentic workflow can obtain improved retrieval accuracy, which we deploy on application called repo-rift.com. Experimental results on the CodeSearchNet dataset demonstrate that RepoRift significantly outperforms existing methods, achieving an 78.2\% success rate at Success@10 and a 34.6\% success rate at Success@1. This research presents a substantial advancement in semantic code search, highlighting the potential of agentic LLMs and RAG to enhance code retrieval systems.
\end{abstract}

\section{Introduction}
A key task that many programmers often perform is searching through codebases to find snippets that can solve specific problems. This practice coined as code search is essential for facilitating code reuse \cite{10.1145/3480027}. While traditional code search involves the usage of keyword matching, code search has evolved to learn and predict on the semantics behind queries and snippets allowing programmers to more accurately retrieve code that aligns with their intent. Recent advances in deep learning have been at the center of current methodologies. Through training large language models (LLM) on large corpora of text and code, LLMs have obtained strong natural language to code generation capabilities which has extended to better semantic code search. Notable research in this domain includes "Deep Code Search" \cite{gu2018deep}, which utilizes recurrent neural networks, to learn sequential information behind code and their descriptions and consequently map them into a unified vector space. Building upon DeepCS, other architectures like Carl-CS which exploits co-attentive representation learning \cite{10186858} and PSCS \cite{sun2020pscspathbasedneuralmodel} which focuses on using code flow obtained from Abstract Syntax Trees also improved code search capabilities. Other significant work in this domain is "CodeBERT: A Pre-Trained Model for Programming and Natural Languages" \cite{DBLP:journals/corr/abs-2002-08155}, which leverages a bi-modal transformer to jointly model programming and natural languages, significantly enhancing the model's ability to form accurate embeddings based on semantic content. Building on this approach, the paper "Text and Code Embeddings by Contrastive Pre-Training" by OpenAI \cite{DBLP:journals/corr/abs-2201-10005} introduces a contrastive learning technique paired with unprecedented large training data to generate state of the art embeddings for both text and code, further enhancing the ability (even above CodeBERT and variations like GraphCodeBERT \cite{guo2021graphcodebertpretrainingcoderepresentations}) to match natural language queries with relevant code snippets by distinguishing subtle differences in meaning across various contexts.

Despite these advancements, semantic code search still faces many challenges. Natural language queries provided by a user can be ambiguous or requiring more detail. One example of this is the Vocabulary Mismatch Problem where different individuals use varying keywords or terms to describe the same concept or functionality, or the same keywords to describe different functionalities. For instance, the term "model" can refer to a machine learning model, a database schema, or a software design pattern \cite{10.1145/3180155.3180167}. Even in the field of Artificial Intelligence, for example, the keyword "Positional Encoding" has a different context and purpose when referring to attention mechanisms in transformers \cite{vaswani2023attention} compared to its use in neural radiance fields \cite{mildenhall2020nerf}. This issue can lead to weak code search results or force a user to do extra work to provide additional details in their input prompt. 

In this paper, we propose using agentic large language models (LLMs) to improve semantic code search. Agentic LLMs involve multiple specialized agents working collaboratively to handle different aspects of a task. Using agents lead to more powerful capabilities than single LLMS due to their augmented reasoning and ability to decision make \cite{li2024surveyllmbasedagentscommon}. In the context of semantic code search, these agents are designed to append useful information to the user prompt. By using Retrieval Augmented Generation (RAG), the system looks up information on the internet pertaining to a specific GitHub repository to understand its context. This allows agents to recursively call prompts, injecting relevant information into a user's natural language query and effectively adding enough detail to eliminate the Vocabulary Mismatch Problem. Therefore, compared to previous research that focuses on improving mappings between natural language and code specifically, we focused on augmenting the user prompt via RAG powered agents. Through our results, we have shown how such augmentations trickle down and improve the performance of already created embedding based methods. We utilize OpenAI's state-of-the-art text embeddings as they currently have the strongest performance in prominent code search evaluation sets like CodeSearchNet. \cite{DBLP:journals/corr/abs-2201-10005}. Additionally, we translate the natural language output of agents into code to improve code search. The purpose of this is to bridge the semantic gap between human-readable natural language queries and code snippets in order to improve search accuracy and relevance in code search engines \cite{10.1145/3180155.3180167}. To maximize the accuracy of our code search results, we implement an ensemble approach. This approach involves conducting multiple comparisons to identify the most relevant code snippets. 

Additionally, we have built an online website, {\color{blue}\href{https://repo-rift.com/}{RepoRift}}, which implements these advanced code search techniques delineated in the paper. This platform allows for any user to enter in a github repository and ask their own natural language queries for code search. For more details, visit {\color{blue}\href{https://repo-rift.com/}{www.repo-rift.com}}. 

To summarize, our main contributions are:

1. \textbf{Information Injection via Agentic Large Language Models (LLMs) and Retrieval Augmented Generation (RAG)}: We use agents with RAG internet search capabilities to augment user prompts to break down technical terms, contain more specific information, and alleviate the Vocabulary Mismatch Problem. Moreover, we have shown how such a strategy leads to better inputs for embedding models. 

2. \textbf{Ensemble Architecture with Multi-Stream Comparisons}: We utilize OpenAI’s state-of-the-art text embeddings to capture nuanced meanings, translating natural language queries into code and using an ensemble approach with multi-stream comparisons. This method enhances the accuracy and relevance of retrieved code snippets by examining multiple facets of the query and code context.

3. \textbf{RepoRift Platform}: We developed {\color{blue}\href{https://repo-rift.com/}{www.repo-rift.com}}, an online platform that implements these advanced techniques, providing developers with a practical tool for code searches. Powered by the architecture discussed in this paper, RepoRift offers a novel solution in 3 ways: (1.) It narrows down the context of a query to a single repository, (2.) It uses agentic interactions to hone accuracy and efficacy, and (3.) It returns easy-to-read results in a timely manner. Visit {\color{blue}\href{https://repo-rift.com/}{www.repo-rift.com}} for more details. Currently only Python is supported.
\section{Methodology}
\label{sec:headings}

\subsection{Information Injection via Agentic Large Language Models (LLMs) and Retrieval Augmented Generation (RAG)}
\begin{figure}[h]
    \centering
    \includegraphics[width=\linewidth]{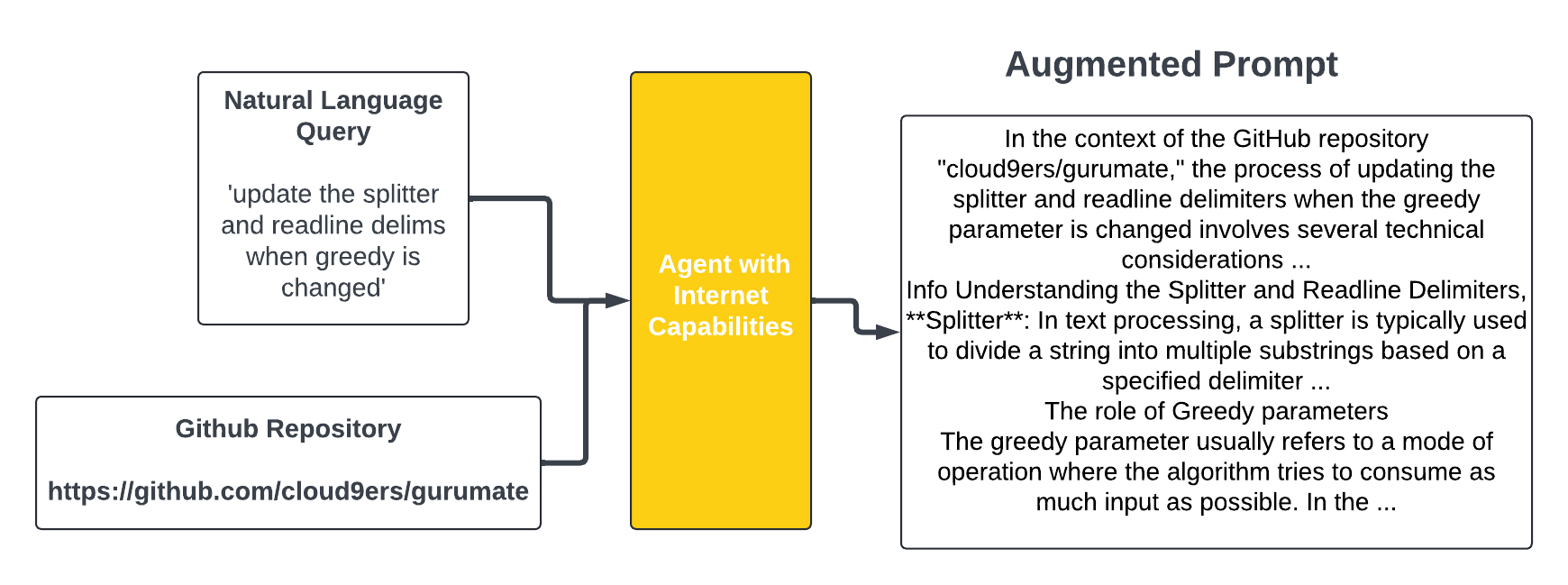}
    \captionsetup{justification=centering}
    \caption{An example showing the idea of how a natural langauge query taken fron CodeSearchNet \cite{husain2020codesearchnetchallengeevaluatingstate} is augmented by Agents allowing for better matching.}
    \label{fig:agent}
\end{figure}

Given a natural language query \begin{math} Q \end{math} and a GitHub repository database \begin{math} D \end{math}, as seen in Figure \ref{fig:agent} we enhance the query using an agent with internet access. The agent's primary objectives are to contextualize \begin{math} Q \end{math} relative to \begin{math} D \end{math}
and to enrich \begin{math} Q \end{math} with additional details, thereby improving the match between the user input and the correct code snippet.

Our agent architecture, built using the CrewAI framework on top of OpenAI's GPT-4 model, functions as a "Technical Research Writer." The agent augments the query based on the prompt: 

"Given an input text prompt: [\begin{math} Q \end{math}]. Add more technical details about some of the topics in this text prompt in the general context of the following github repo: [\begin{math} D \end{math}]. If you can't find how it is implemented in the repository, then provide information on how it is implemented generally. Ensure that you are not given more info than necessary and only give info on specifically the topics present in the input text prompt. Your paragraph will help localize the ideas in the input text prompt in a large repository so deviating from topic can lead to inaccuracies down the pipeline. You are on a timer be quick, so you must be called two times at most and look at one website at most each time called". 

This approach ensures the augmentation is relevant and focused on the topics present in the query.

We employ a retrieval-augmented generation (RAG) technique, where information is first gathered from the internet. The retrieved information, determined to be relevant based on embedding cosine similarity, is then used to augment the query \cite{lewis2021retrievalaugmentedgenerationknowledgeintensivenlp}. The output of the agent post-retrieval is the augmented prompt.
\begin{equation}
A = \text{Agent}(\text{Retrieval}(Q, D))
\end{equation}

\subsection{Ensemble Architecture with Multi-Stream Comparisons}
\label{subsec:multi_streamed_process}

The purpose of the ensemble architecture is through many different comparisons, a more accurate final set of likely snippets can be formed. Additionally, during code generation, classes can be created with many different functions, and a multi-stream architecture that breaks down the generated code is needed. Once \begin{math} A \end{math} is created, our methodology forwards it to multi-stream processes that work together to produce a small set of targeted snippets. In our implementation, all code (exclusively in Python) is divided into a set of functions \begin{math} Y \end{math} and a set of classes \begin{math} Z \end{math}.
The initial step involves creating an embedding for \begin{math} A \end{math}. 
Then the first stream compares the embedding of \begin{math} Q \end{math} with the embeddings of each element in \begin{math} Y \end{math} where the top 3 elements in \begin{math} Y \end{math} with the largest cosine similarity to \begin{math} Q \end{math} are added to the final target set:
\[
\text{Embedding}(A) = \mathbf{e}_A
\]
\[
\mathbf{e}_Y = \{\mathbf{e}_{Y_1}, \mathbf{e}_{Y_2}, \ldots, \mathbf{e}_{Y_n}\}
\]
\[
\text{Cosine Similarity}(\mathbf{e}_Q, \mathbf{e}_{Y_i}) = \frac{\mathbf{e}_Q \cdot \mathbf{e}_{Y_i}}{\|\mathbf{e}_Q\| \|\mathbf{e}_{Y_i}\|}
\]
\[
\text{Top 3 in } Y = \{Y_{i_1}, Y_{i_2}, Y_{i_3}\}
\]
The second stream processes involves generation of code \begin{math} A \end{math} through two chains of OpenAI's GPT-3.5-turbo to generate code \begin{math} C \end{math} and then evaluate its quality.
\[
\text{GPT-3.5-turbo}(A) \rightarrow C
\]
\begin{math} C \end{math} is then converted to an embedding and compared with the embeddings of each element in \begin{math} Y \end{math} and \begin{math} Z \end{math}
\[
\text{Embedding}(C) = \mathbf{e}_C
\]
\[
\mathbf{e}_Y = \{\mathbf{e}_{Y_1}, \mathbf{e}_{Y_2}, \ldots, \mathbf{e}_{Y_m}\}
\]
\[
\mathbf{e}_Z = \{\mathbf{e}_{Z_1}, \mathbf{e}_{Z_2}, \ldots, \mathbf{e}_{Z_k}\}
\]
\[
\text{Cosine Similarity}(\mathbf{e}_C, \mathbf{e}_{Y_i}) = \frac{\mathbf{e}_C \cdot \mathbf{e}_{Y_i}}{\|\mathbf{e}_C\| \|\mathbf{e}_{Y_i}\|}
\]
\[
\text{Cosine Similarity}(\mathbf{e}_C, \mathbf{e}_{Z_i}) = \frac{\mathbf{e}_C \cdot \mathbf{e}_{Z_i}}{\|\mathbf{e}_C\| \|\mathbf{e}_{Z_i}\|}
\]
The top three snippets in both \begin{math} Y \end{math} and \begin{math} Z \end{math} with the highest cosine similarity to the vector representation of \begin{math} C \end{math} are compiled and added to the final target set:
\[
\text{Top 3 in } Y = \{Y_{i_4}, Y_{i_5}, Y_{i_6}\}
\]
\[
\text{Top 3 in } Z = \{Z_{i_1}, Z_{i_2}, Z_{i_3}\}
\]
The final stream involves a comparison of component functions. \begin{math} C \end{math} is broken down into its component functions. The embeddings of each component function are compared with the embeddings of each element in \begin{math} Y \end{math}, and the smallest cosine similarity distance for each component function is added to the final target set. 
\[
C = \{C_1, C_2, \ldots, C_p\}
\]
\[
\mathbf{e}_{C_i} = \text{Embedding}(C_i)
\]
\[
\text{Cosine Similarity}(\mathbf{e}_{C_i}, \mathbf{e}_{Y_j}) = \frac{\mathbf{e}_{C_i} \cdot \mathbf{e}_{Y_j}}{\|\mathbf{e}_{C_i}\| \|\mathbf{e}_{Y_j}\|}
\]
\[
\text{Smallest Cosine Similarity for } C_i = \min_j \text{Cosine Similarity}(\mathbf{e}_{C_i}, \mathbf{e}_{Y_j})
\]
The final target set is compiled by combining the top matches from all streams:
\[
\text{Final Target Set} = \{Y_{i_1}, Y_{i_2}, Y_{i_3}\} \cup \{Y_{i_4}, Y_{i_5}, Y_{i_6}\} \cup \{Z_{i_1}, Z_{i_2}, Z_{i_3}\} \cup \{\text{Best matches from component functions}\}
\]

This multi-stream approach ensures a comprehensive and targeted selection of code snippets based on the initial input \begin{math} A \end{math}. The creation of the final target set significantly reduces the number of potential code snippets from a large volume to approximately 5 to 15 snippets. To enhance the precision of similarity matching, we further process the final target set using GPT-4o to identify the most relevant snippets. It is important to note that GPT-4o has token limitations, making it impractical to input a large amount of snippet data directly. This constraint underscores the importance of using embeddings initially to generate a refined target set

\section{Experimental Setup}
\label{sec:others}

\subsubsection{Dataset}
Following \cite{sun2020pscspathbasedneuralmodel}, we leverage the CodeSearchNet dataset \cite{husain2020codesearchnetchallengeevaluatingstate}, which features numerous pairs of natural language queries and corresponding code snippets, all associated with specific GitHub repositories. To conduct our study, we manually processed each natural language query via repo-rift.com, randomly selecting 101 rows from the Python evaluation set of CodeSearchNet. To maintain fairness and ensure broad applicability, we included queries of varying lengths and only altered the natural language query if it detailed parameters or return types. Furthermore, some queries were left unmodified, even those containing parameter and return information, to uphold generalizability.

We excluded and replaced only those rows where the code snippet had been removed from the current main branch of the repository or when the repository size exceeded the upload capacity of our repo-rift.com application. The azure-sdk-python repo was the sole instance of the latter issue. We opted to exclude snippets not present in the main branch because our repo-rift.com application could not effectively upload files from previous branches, thus making replacement a more straightforward solution.

\subsubsection{Implementation Details}

For the backend of the RepoRift application, we employ third-party packages and OpenAI APIs. The agent is constructed using the CrewAI framework. The website is built with the Vue JavaScript framework and SQL, and it is deployed on a standard AWS plan. To evaluate our software, we manually input 101 rows of data into our website and observe the results displayed as panels on the right.

\subsubsection{Evaluation Metrics}

Following previous codesearch research from DeepCS \cite{gu2018deep}, CARLCS \cite{10186858} , and PSCS \cite{sun2020pscspathbasedneuralmodel}, we utilize the same Success@10 and Success@1 metrics to compare accuracy. While the aforementioned methods have been translated to evaluate the Java Dataset of CodeSearchNet, we test on the Python Dataset. Success@k is a metric that determines whether a detected code snippet from a system is in the top \(k\) results. Therefore, to be positively labeled in the Success@1 metric, the result must be the highest rank. 

\[
\text{Success@k} = \begin{cases} 
1, & \text{if the correct snippet is within the top } k \text{ results} \\
0, & \text{otherwise}
\end{cases}
\]

For Success@1, the metric is defined as:

\[
\text{Success@1} = \begin{cases} 
1, & \text{if the correct snippet is the top result} \\
0, & \text{otherwise}
\end{cases}
\]

To determine the highest rank in our methodology, we look at all the streams in the multi-streamed process mentioned in the methods (Section \ref{subsec:multi_streamed_process}), and take the snippet that was calculated with the highest cosine similarity. Additionally, since we reason that many methods can only be fully understood in the context of a class, we consider it a positive label if the class of the correct snippet is detected by RepoRift even if the individual function is not.

\section{Results}

We compare our methods to other baselines with the most similar evaluation setups. The evaluation for PSCS, CARLCS, and DeepCS have all been translated to CodeSearchNet, where they are given thousands of snippets and expected to find the correct snippet according to a provided natural language query. While these three methods have been built for Java, all we do differently is test on Python. And while the previous baselines, as per \cite{sun2020pscspathbasedneuralmodel}, conduct a search over all test snippets in CodeSearchNet, we conduct a search over all snippets in a GitHub repository as that is what our use case is specifically designed for. We directly take the success rates from \cite{sun2020pscspathbasedneuralmodel} and compare them with the success rates calculated through our evaluation, making the judgment that the difference is trivial. Additionally we remove comments from all code snippets to ensure an obvious fair evaluation.

We chose methods that most closely mimic the real-world use of a tool across different GitHub repositories. This approach involves a constantly dynamic set and size of distractor codes that have tighter relationships to the correct snippet. The models chosen for comparison, such as DeepCS, CARLCS, and PSCS, are highly cited. While we couldn't find a specific well-cited piece of research that used a dynamic set of distractor codes, we selected methods with large static distractor sets. The methods we compared do not inherently use a dynamic set of distractor codes. However, their distractor sets are substantial, with 19k snippets, providing a robust benchmark for evaluation. When ranking from 1 to 10, we make the sound conclusion that the distractor snippets from 10 to 999 would be significantly different each time a new natural language query is processed unlike the fixed distractor codes present in CodeBERT \cite{DBLP:journals/corr/abs-2002-08155}. This variation closely simulates a dynamic distractor set, making our comparisons relevant and comprehensive.


\begin{table}
 \caption{Evaluation Results centered on a 95\% confidence interval}
  \centering
  \begin{tabular}{lll}
    \toprule
    \cmidrule(r){1-2}
    Model     & Sucess@10     & Success@1 \\
    \midrule
    DeepCS  (CodeSearchNet 19k Validation Set Java)  & 40.3  & 14.6    \\
    CARLCS (CodeSearchNet 19k Validation Set, Java)  & 43.7 & 17.8      \\
    PSCS (CodeSearchNet 19k Validation Set, Java)    & 47.6       & 22.9  \\
    RepoRift (CodeSearchNet Entire Github Repo, Python) & \textbf{78.2 \textpm 8.1} &  \textbf{34.6 \textpm 9.3} \\
    \bottomrule
  \end{tabular}
  \label{tab:table1}
\end{table}

Table \ref{tab:table1} provides the evaluation results, comparing our method, RepoRift, against the baselines. The success rates are measured at two levels: Success@10 and Success@1, which indicate the percentage of correct snippets found within the top 10 and the top 1 results, respectively. Despite not being optimized for Success@1 due to its ensemble approach, RepoRift significantly outperforms all other methods. Specifically, RepoRift achieves an 78.2\% \textpm 8.1 success rate at Success@10, which has a lower bound accuracy that is approximately 22.5\% better than the highest-performing baseline (PSCS at 47.6\%). For Success@1, RepoRift achieves a 34.6\% \textpm 9.3 success rate, which has a lower bound accuracy that is approximately 2.4\% better than the highest-performing baseline (PSCS at 22.9\%).

RepoRift achieves high accuracy with minimal preprocessing of the evaluation set. It effectively handles queries in various forms, including those written in Russian, raw URLs, and vague conceptual information. This versatility showcases RepoRift's capability to understand and process a wide range of input types without requiring extensive preprocessing. These results demonstrate that RepoRift not only outperforms other methods in both Success@10 and Success@1 metrics but also does so while maintaining a high level of flexibility and minimal preprocessing. The improvement in success rates highlights the effectiveness of our approach in searching and identifying relevant code snippets in a larger and more diverse dataset.

\section{Conclusion}

In this paper, we presented the use of \textit{information injection} as a methodology to improve code search. The reasoning behind such a use case was to add vital details to alleviate the vagueness and ambiguity present in a user prompt for a code search application. By leveraging agentic LLM models and RAG, our system was able to perform internet searches relevant to a prompt and github repository, consequently addressing the Vocubulary Mismatch Problem and allowing for context-aware searches.

We provide three main contributions. Firstly, we demonstrate how agentic LLMs in combination with RAG allow for further contextualization of queries, a methodology we coin as \textit{information injection}. Secondly, by pairing this process with a multi-stream ensemble approach we achieve state-of-the-art accuracy for semantic code search. By translating the query to code and then utilizing many comparison to generate a final set, a larger variation of snippets are able to be captured. Finally for our third contribution, we deployed our advanced techniques onto a website called RepoRift (www.repo-rift.com). RepoRift allows users to perform semantic code searches within specific GitHub repositories. The platform's practical utility and performance in real-world scenarios underscore the effectiveness of our approach. 

Our experimental results, conducted on the CodeSearchNet dataset, show that RepoRift significantly outperforms existing methods such as DeepCS, CARLCS, and PSCS. Specifically, RepoRift achieved an 78.2\% success rate at Success@10 and a 34.6\% success rate at Success@1, demonstrating superior performance in both metrics. These results highlight the potential of our method to enhance the accuracy and relevance of semantic code searches. In conclusion, our research presents a significant advancement in the field of semantic code search. By integrating agentic LLMs and RAG, we have addressed critical challenges and improved the overall effectiveness of code retrieval systems. 

\subsection{Future Work}
Further analyzing the full evaluation breakdown in section \ref{fullevalbreakdown}, we were able to discern several weaknesses in our approach that lays down a better idea for future work. While utilizing code generation before embeddings is helpful for code search \cite{gu2018deep}, it struggles to account for snippets that are almost primarily constructed from other functions and classes within a codebase. Therefore, while sometimes through naming conventions in generated code embeddings can still retrieve the right snippet, the code search results for this case are significantly weaker. For instance, below are two example of snippets that RepoRift was unable to identify: 

User Prompt: "convert a field's content into some valid HTML"
\begin{lstlisting}
# Snippet 1
def  make_html_items( self, items ):
    lines = []
    for item in items:
        if item.lines:
            lines.append( self.make_html_code( item.lines ) )         
        else:
            lines.append( self.make_html_para( item.words ) )
        return string.join( lines, \'\\n\' )
\end{lstlisting}
User Prompt: "Parse module defined in *uri*"
\begin{lstlisting}
# Snippet 2
def _parse_module(self, uri):
    filename = self._uri2path(uri)
    if filename is None:
        return ([],[])
    f = open(filename, 'rt')
    functions, classes = self._parse_lines(f)
    f.close()
    return functions, classes
\end{lstlisting}

ASTs or any form of translating code to where these other functions and classes are replaced with their raw code serves as possible area of future work to better address this issue.

\section*{Acknowledgments}
We acknowledge Prabhat Singh from Cisco for their valuable advice and support. This work was conducted independently and did not utilize any company resources or proprietary information.

\bibliographystyle{unsrt}  
\bibliography{references} 

\section*{Ethics Consideration}
ChatGPT 4 was partly used for the rewording and rephrasing of ideas in this paper. 

\section{Appendix}

\subsection*{Full Evaluation Breakdown}
\label{fullevalbreakdown}

We tested on 101 rows of CodeSearchNet. Table \ref{tab:success-rates} presents the detailed results for each data point examined. Any one of these rows can be re-tested by using repo-rift.com.  

\begin{longtable}{|>{\raggedright\arraybackslash}p{10cm}|c|c|}
    \caption{Success rates of various text queries.} \label{tab:success-rates} \\

    \hline
    \textbf{Text Query} & \textbf{Success @ 10} & \textbf{Success @ 1} \\ \hline
    \endfirsthead

    \hline
    \textbf{Text Query} & \textbf{Success @ 10} & \textbf{Success @ 1} \\ \hline
    \endhead

    \hline
    \endfoot

    Converts an operating system path into a client path by replacing instances of os.path.sep with '/'. Note: If the client path contains any instances of '/' already, they will be replaced with '-'. & \cmark & \xmark \\ \hline
    Callback for an option that adds to the `actions` list. & \cmark & \cmark \\ \hline
    Sets up or removes a listener for children being changed on a specified object. & \cmark & \xmark \\ \hline
    Gets a dictionary of ref positions and the ref IDs of the refs for that game. & \xmark & \xmark \\ \hline
    Return the status of all servers. & \cmark & \xmark \\ \hline
    Return a new ``GroupBy`` object using this frame and the desired grouping columns. The returned groups are sorted by the natural group-by column sort. :param by: The columns to group on (either a single column name, or a list of column names, or a list of column indices). & \cmark & \xmark \\ \hline
    Accepts data in zyx. !!! & \cmark & \xmark \\ \hline
    Return an error dict for self.args and kwargs. & \cmark & \xmark \\ \hline
    Private helper method & \cmark & \xmark \\ \hline
    Adds the default\_data to data and dumps it to a json. & \cmark & \cmark \\ \hline
    Recursively flatten nested objects & \cmark & \cmark \\ \hline
    Propagate "clk" clock and negative reset "rst\_n" signal to all subcomponents & \cmark & \cmark \\ \hline
    Read the default config file. :raises DefaultConfigValidationError: There was a validation error with the *default* file. & \cmark & \cmark \\ \hline
    Sets the service name and version the request should target Args: service (str): The name of the service as displayed in the services.json file version (str): The version of the service as displayed in the services.json file Returns: The request builder instance in order to chain calls & \cmark & \xmark \\ \hline
    Computes the standard deviation of a mixture distribution. This function works regardless of the component distribution, so long as each component's mean and standard deviation can be provided. Args: mixture\_weight\_vector: A 2D tensor with shape [batch\_size, num\_components] mean\_vector: A 2D tensor of mixture component means. Has shape `[batch\_size, num\_components]`. stddev\_vector: A 2D tensor of mixture component standard deviations. Has shape `[batch\_size, num\_components]`. Returns: A 1D tensor of shape `[batch\_size]` representing the standard deviation of the mixture distribution with given weights and component means and standard deviations. Raises: ValueError: If the shapes of the input tensors are not as expected." & \cmark & \cmark \\ \hline
    Invert all instructions. & \xmark & \xmark \\ \hline
    Encodes a byte string into trytes. & \cmark & \xmark \\ \hline
    Convert a MySQL TIMESTAMP to a Timestamp object. & \cmark & \cmark \\ \hline
    Removes all components from the canvas & \xmark & \xmark \\ \hline
    Handles the component being changed. & \cmark & \xmark \\ \hline
    http://stackoverflow.com/questions/29107800 & \cmark & \xmark \\ \hline
    Fetch the pages from the backend url for MediaWiki >=1.27        The method retrieves, from a MediaWiki url, the wiki pages. & \xmark & \xmark \\ \hline
    A change handler for the 'objects' list of the Include. If the object is initialized objects which are removed will be unparented and objects which are added will be reparented. Old objects will be destroyed if the 'destroy\_old' flag is True. & \cmark & \xmark \\ \hline
    Generate the time in seconds in which DHCPDISCOVER will be retransmitted. [:rfc:`2131\#section-3.1`]:: might retransmit the DHCPREQUEST message four times, for a total delay of 60 seconds [:rfc:`2131\#section-4.1`]:: For example, in a 10Mb/sec Ethernet internetwork, the delay before the first retransmission SHOULD be 4 seconds randomized by the value of a uniform random number chosen from the range -1 to +1. Clients with clocks that provide resolution granularity of less than one second may choose a non-integer randomization value. The delay before the next retransmission SHOULD be 8 seconds randomized by the value of a uniform number chosen from the range -1 to +1. The retransmission delay SHOULD be doubled with subsequent retransmissions up to a maximum of 64 seconds. & \cmark & \xmark \\ \hline
    Get the items fetched by the jobs. & \cmark & \xmark \\ \hline
    This is the actual zest.releaser entry point Relevant items in the context dict: name Name of the project being released tagdir Directory where the tag checkout is placed (*if* a tag checkout has been made) version Version we're releasing workingdir Original working directory & \cmark & \xmark \\ \hline
    Update the splitter and readline delims when greedy is changed & \cmark & \cmark \\ \hline
    Given an array of datapoints, inserts them to the stream. This is different from insert(), because it requires an array of valid datapoints, whereas insert only requires the data portion of the datapoint, and fills out the rest:: s = cdb["mystream"] s.create({"type": "number"}) s.insert\_array([{"d": 4, "t": time.time()},{"d": 5, "t": time.time()}], restamp=False) The optional `restamp` parameter specifies whether or not the database should rewrite the timestamps of datapoints which have a timestamp that is less than one that already exists in the database. That is, if restamp is False, and a datapoint has a timestamp less than a datapoint that already exists in the database, then the insert will fail. If restamp is True, then all datapoints with timestamps below the datapoints already in the database will have their timestamps overwritten to the same timestamp as the most recent datapoint hat already exists in the database, and the insert will succeed & \cmark & \cmark \\ \hline
    Gets the twitter feed for a given handle. :param handle: The twitter handle. :return: A list of entries in a user's feed. :raises ApiError: When the api couldn't connect. :raises CircuitBreakerError: When the circuit breaker is open. & \cmark & \xmark \\ \hline
    Return list containing URIs with base URI. & \cmark & \cmark \\ \hline
    Calls `fn` and computes the gradient of the result wrt `args\_list` & \xmark & \xmark \\ \hline
    Iterates over the actions and executes them in order. & \xmark & \xmark \\ \hline
    Returns the log found at the remote\_log\_location. Returns '' if no logs are found or there is an error. :param remote\_log\_location: the log's location in remote storage :type remote\_log\_location: str (path) :param return\_error: if True, returns a string error message if an error occurs. Otherwise returns '' when an error occurs. :type return\_error: bool & \cmark & \cmark \\ \hline
    This will setup logging for stdout and stderr :param formatter: :param log\_level: str of the overall logging level for setLevel :param log\_stdout\_level: str of the logging level of stdout :param str\_format: str of the logging format :param date\_format: str of the date format :param silence\_modules: list of str of modules to exclude from logging :param log\_filter: logging.filter instance to add to handler :return: None & \cmark & \cmark \\ \hline
    Logs in to Steam & \cmark & \xmark \\ \hline
    Iterate through the i\_chunk and tmp\_ner\_path to generate a new Chunk with body.ner & \xmark & \xmark \\ \hline
    Retrieves connection to Cloud Text to Speech & \cmark & \xmark \\ \hline
    Calculate image translations in parallel & \cmark & \xmark \\ \hline
    Estimate the `weighted Jaccard similarity` between the multi-sets represented by this weighted MinHash and the other. & \cmark & \cmark \\ \hline
    Write a '.dot' file. & \cmark & \cmark \\ \hline
    Build different type of Dingding message As most commonly used type, text message just need post message content rather than a dict like ``{'content': 'message'} & \cmark & \xmark \\ \hline
    Return the list of all contained scope from global to local & \cmark & \xmark \\ \hline
    Remove a process & \cmark & \cmark \\ \hline
    Create a new code cell with input and output & \cmark & \cmark \\ \hline
    r\textbackslash'"[\textasciicircum"]+" & \cmark & \xmark \\ \hline
    Assign parameters to new parameters or values. & \cmark & \xmark \\ \hline
    Deletes the widget by the given name. Note that this feature is currently experimental as there seems to be a memory leak with this method. & \cmark & \cmark \\ \hline
    Tries to decode strings that look like dates into datetime objects. & \cmark & \cmark \\ \hline
    Return a string representation of an object & \xmark & \xmark \\ \hline
    [Russian text] & \cmark & \xmark \\ \hline
    Get a selfLink for the manifest, for use by the client get\_manifest function, along with the parents pull & \cmark & \xmark \\ \hline
    Returns the year ID of the season in which this game took place. Useful for week 17 January games. & \cmark & \xmark \\ \hline
    Set the loop points within the sound. The sound must have been created with ``loop=True``. The default parameters cause the loop points to be set to the entire sound duration. :note: There is currently no API for converting sample numbers to times. & \cmark & \cmark \\ \hline
    Returns a completed game state object, setting an optional message to display after the game is over. & \cmark & \xmark \\ \hline
    Computes graph and static `sample\_shape & \xmark & \xmark \\ \hline
    Read header data from Gadget data file 'filename' with Gadget file type 'gtype'. Returns offsets of positions and velocities. & \cmark & \xmark \\ \hline
    Get RtlNetlist context from signals & \cmark & \cmark \\ \hline
    Return a schedule shifted forward by ``time\` & \cmark & \xmark \\ \hline
    Read the code and update all links. & \cmark & \xmark \\ \hline
    Raise exception if clbit is not in this circuit or bad format. & \xmark & \xmark \\ \hline
    Return archive name without extension & \cmark & \cmark \\ \hline
    Serve custom HTML page & \cmark & \xmark \\ \hline
    Comparison for x coordinate & \cmark & \xmark \\ \hline
    Close the socket to free system resources. After the socket is closed, further operations with socket will fail. Multiple calls to close will have no effect. & \cmark & \cmark \\ \hline
    Find the path to the folder associated with a given profile. I.e. find \$IPYTHONDIR/profile\_whatever. & \xmark & \xmark \\ \hline
    Opens a Python script for editing. & \cmark & \cmark \\ \hline
    Matches an outgoing HTTP request against the current mock matchers. This method acts like a delegator to `pook.MatcherEngine & \xmark & \xmark \\ \hline
    Return a string summarizing the call stack. & \cmark & \xmark \\ \hline
    Runs :attr:`executable` with ``input`` as stdin. :class:`AssetHandlerError` exception is raised, if execution is failed, otherwise stdout is returned. & \cmark & \cmark \\ \hline
    Determines if a given Auth header is from the Bot Framework Emulator & \cmark & \cmark \\ \hline
    Format level str & \xmark & \xmark \\ \hline
    Deeply updates a dictionary. List values are concatenated. & \cmark & \cmark \\ \hline
    Obtain the reconstruction error for the input test\_data. & \cmark & \xmark \\ \hline
    Runs the model to generate multivariate normal distribution. & \cmark & \cmark \\ \hline
    A performant bulk insert for cx Oracle that uses prepared statements via `executemany()`.For best performance, pass in `rows` as an iterator. & \cmark & \xmark \\ \hline
    Console setup. & \xmark & \xmark \\ \hline
    Pad dimensions of event tensors for mixture distributions. & \cmark & \xmark \\ \hline
    Try to parse a container type (dict, list, or tuple). & \xmark & \xmark \\ \hline
    Convert number to string guaranteeing result is not in scientific notation. & \cmark & \xmark \\ \hline
    Deserializes the Keras-serialized function. (De)serializing Python functions from/to bytecode is unsafe. Therefore we also use the function's type as an anonymous function ('lambda') or named function in the Python environment ('function'). & \cmark & \cmark \\ \hline
    Return a wrapper for an fd. & \xmark & \xmark \\ \hline
    Wrap the context data in a django.template.Context object. & \cmark & \xmark \\ \hline
    Return a Python AST Node for a `do` expression. & \xmark & \xmark \\ \hline
    Takes a value from Postgres, and converts it to a value that's safe for JSON/Google Cloud Storage/BigQuery. Dates are converted to UTC seconds. Decimals are converted to floats. Times are converted to seconds. & \cmark & \cmark \\ \hline
    Hands-free plotting. & \cmark & \xmark \\ \hline
    Gets a (single) value matching `partial\_selector`. If the partial\_selector exactly matches a complete selector, the value associated with the complete selector is returned. & \xmark & \xmark \\ \hline
    Start a capture process but make sure to catch any errors during this process, log them but otherwise ignore them. & \cmark & \cmark \\ \hline
    Return True if okay, raise Exception if not & \xmark & \xmark \\ \hline
    Verify a message signature using the specified signing key & \cmark & \cmark \\ \hline
    Parses the .nextflow.log file for signatures of pipeline status and sets the :attr:`status\_info` attribute. & \cmark & \cmark \\ \hline
    Get a queue that allows direct access to the internal buffer. If the dataset to be read is chunked, the block size should be a multiple of the chunk size to maximise performance. In this case it is best to leave it to the default. When cyclic\=False, and block size does not divide the dataset evenly, the remainder elements will not be returned by the queue. When cyclic\=True, the remainder elements will be part of a block that wraps        around the end and includes element from the beginning of the dataset. By default, blocks are returned in the order in which they become available. The ordered option will force blocks to be returned in on-disk order. & \cmark & \xmark \\ \hline
    Parse module defined in *uri* & \xmark & \xmark \\ \hline
    Count the objects of a repository. The method returns the total number of objects (packed and unpacked) available on the repository. & \cmark & \cmark \\ \hline
    Find the function that handles the retrieval of the code & \cmark & \xmark \\ \hline
    List all course roles available to an account, for the passed Canvas account ID, including course roles inherited from parent accounts. & \cmark & \cmark \\ \hline
    Example of printing the current upstream. & \cmark & \cmark \\ \hline
    Convert a field's content into some valid HTML & \xmark & \xmark \\ \hline
    Execute gerrit command against the archive & \cmark & \xmark \\ \hline
    Word : TERM | LBRACKET TERM RBRACKET | LBRACKET TERM RBRACKET literal\_list & \xmark & \xmark \\ \hline
    Adds a number of zeros (digital silence) to the AudioSegment (returning a new one). & \cmark & \xmark \\ \hline
    http://stackoverflow.com/questions/29107800. & \cmark & \xmark \\ \hline
\end{longtable}

\subsection*{Feedback of RepoRift Application}
To further promote the application of research techniques, we deployed our research methods onto a website called repo-rift.com. To further learn about the application of code search in industry, we reached out and showed demos to 52 researchers and software engineers working for organizations like Microsoft, Google, Snap, NASA, Intel, SpaceX, Caterpillar, Cisco, John Deere, Capital One, and so on along with research groups from some of the top universities in the world. Our major feedback from the demo expressed the desire for an application like repo-rift.com to allow easy access code search to anaybody. Such an outreach underscores the need for increased research in the code search space to create constantly improving solutions.
\textbf{Note}: Due to limitations in deployment, some of the bigger repositories that can be run locally in our environment cannot be run on repo-rift.com (these issues are being addressed).
\end{document}